\documentclass[%
aps,%
prl,%
twocolumn,
showpacs,%
preprintnumbers,%
byrevtex,%
floatfix%
]{revtex4}

\usepackage{epsf}
\usepackage{ifthen}
\usepackage[usenames]{color}
\usepackage{amssymb,mathbbol,amsmath,amsbsy}
\usepackage{amsfonts}
\usepackage{dsfont}
\usepackage{amsmath}
\usepackage[dvips]{graphicx}
\usepackage[dvips, colorlinks=true, bookmarks, pdftitle={An example PDF file from LaTeX},
 pdfauthor={Gianaurelio Cuniberti}, pdfsubject={From LaTeX to PDF}, pdfkeywords={PDF, LaTeX, hyperlinks, hyperref}]{hyperref}
\def\ii{\textrm{i}\,}

\def\eg{\textit{e.g.}}

\begin{document}
\date{June, 05 2009}
\title{Engineering the thermopower in semiconductor-molecule junctions: towards high thermoelectric efficiency at the nanoscale}
\author{D. Nozaki}
\author{H. Sevin{\c{c}}li}
\author{W. Li}
\author{R.~Guti{\'e}rrez}
\author{G.~Cuniberti}
\affiliation{%
Institute for Materials Science and Max Bergmann Center of Biomaterials,\\ Dresden University of Technology, 01062 Dresden, Germany
}

\begin{abstract}
We propose a possible route to achieve high thermoelectric efficiency in molecular junctions by combining a local chemical tuning of the molecular electronic states with the use of semiconducting electrodes. The former allows to control the position of the HOMO transmission resonance with respect to the Fermi energy while the latter fulfills a twofold purpose: the suppression of electron-like contributions to the thermopower and the cut-off of  the HOMO transmission tails into the semiconductor band gap. As a result a large thermopower can be obtained. Our results strongly suggest that large figures of merit in such molecular junctions can be achieved.

\end{abstract}


\maketitle

\paragraph{Introduction$-$}Thermoelectric materials convert thermal gradients and electric fields for power conversion and for
refrigeration, respectively. With increasing energy demand, thermoelectric applications are attracting a considerable interest. Unfortunately, thermoelectrics find currently only special applications due  to their
limited efficiency, which is measured by a dimensionless parameter, the thermoelectric figure of
merit: $ZT = S^{2}TG/\kappa$, which includes the Seebeck coefficient (thermopower) $S$, an average temperature $T$, the
electrical conductance $G$, and the thermal conductance $\kappa=\kappa_{\textrm{el}}+\kappa_{\textrm{ph}}$, the latter containing both electronic $\kappa_{\textrm{el}}$ and
vibrational $\kappa_{\textrm{ph}}$ contributions. Maximizing $ZT$ is challenging, because optimizing one physical parameter
often adversely affects another.
 
However, as suggested in the early 90s~\cite{PhysRevB.47.12727}, dimensionality reduction towards the nanoscale could provide an additional parameter to tune the electrical and thermal response of thermoelectric materials. This has triggered an active research on new nanoscaled thermoelectric materials~\cite{Boukail2008,Hochbaum2008,li:3186,nielsch2008,mori2005,MingoN._nl034721i,Mingo2008,galli2008,kim:045901,Lyeo2004}. An alternative to inorganic-based  materials could be to exploit  molecules, which have been already extensively investigated in the context of charge transport and molecular electronic applications in the past two decades~\cite{cfr05}. Indeed, the Seebeck coefficient  of different single molecules~\cite{Reddy2007} as well as the thermal conductance of self-assembled monolayers~\cite{wang2007} have been meanwhile experimentally investigated. Theoretically, only few studies on the thermoelectric properties in few level systems (molecules, small quantum dots) have been presented to date
~\cite{
citeulike:3871696,
citeulike:3500670,
citeulike:4270003,
citeulike:3806071,
PhysRevB.70.195107,
citeulike:4785902,
citeulike:3929829,
baranger:nanolett2009,
esfarjani:prb2006,muller:jchemphys2008}.
Concerning this ''organic route'', two aspects should be taken into account. First, as proposed by Mahan and Sofo~\cite{Mahan1996}, the presence of a sharp resonance near the Fermi level $E_{\textrm{F}}$  can considerably increase the thermopower, since  the latter depends on the derivative of the conductance near E$_{\textrm{F}}$. One key advantage of using molecules as potential thermoelectrics is the capability  to tune their chemical, and hence also their electrical and thermal properties in a {\textit{very controlled}} way.  The second aspect is more generic: by using {\textit{metallic}} electrodes to contact a molecule, there are hole-like and electron-like contributions to $S$  arising from tunneling through the highest-occupied molecular orbital (HOMO) and lowest-unoccupied molecular orbital (LUMO) states, respectively. Both contributions appear with different signs in $S$ and thus partially cancel each other. Could one  overcome this problem?

\begin{figure}[b]
\centerline{
\epsfclipon
\includegraphics[width=1.\linewidth]{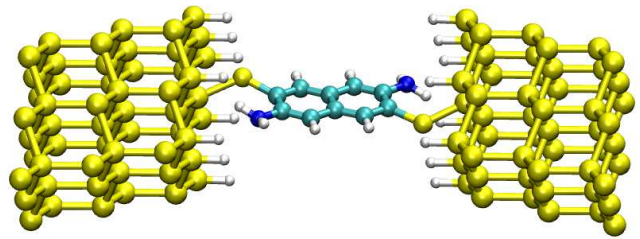}%
}
\caption{\label{fig:fig1}%
A typical molecular junction investigated in this study. Two silicon electrodes terminated at the (111) facet are bridged by a molecule. In the calculations, periodic boundary conditions in the lateral directions have been used. The electrode surface is passivated with hydrogen atoms to avoid strong structural distortions related to surface reconstruction effects. 
}
\end{figure}
In this Letter, we propose  a combination of chemical tuning of the molecular electronic structure with the use of {\textit{semiconducting}} electrodes to improve the thermoelectric efficiency of molecular junctions. Such electrodes provide a way to lift the near cancelation of hole-like and  electron-like contributions to the thermopower. Specifically,  if the LUMO resonance  lies within the semiconductor band gap, no states are available for the ``electronic'' channel, which is then blocked and the thermopower will be mainly determined by the HOMO (hole) channel. This together with the inclusion of donor groups in a $\pi$-conjugated molecule (chemical tuning) can lead to a dramatic increase of the thermopower  of the junction. We illustrate these ideas by using a first-principle based approach to the electronic structure and the transport properties of the molecular junction as well as by a minimal model Hamiltonian. A typical electrode-molecule set-up used in our simulations is displayed in Fig.~\ref{fig:fig1}~\cite{comm2}. 

\begin{figure}[t]
	\begin{center}
	\includegraphics[scale=0.95]{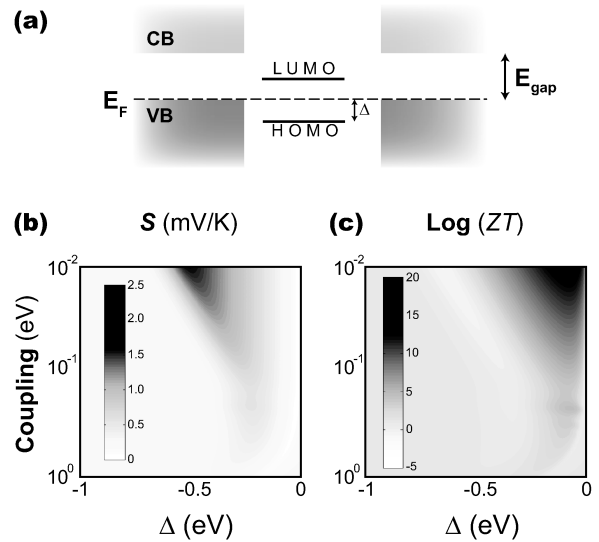}
	\caption{
	Top panel: Energy profile of the model semiconductor-molecule-semiconductor junction.
	$E_{gap}$ is the gap between the valence (VB) and conduction (CB) bands, $\Delta=E_{\textrm{HOMO}}-E_{\textrm{F}}$ denotes the position of the HOMO with respect to $E_{\textrm{F}}$. The molecular gap is chosen so as to have the LUMO level  placed inside the semiconducting gap. Lower panel: Seebeck coefficient $S$ (b)  and thermoelectric figure of merit $\textrm(ZT)$ (c) as a function of $\Delta$ and the effective coupling strength to the electrodes $\Gamma$ at $T=300$ K. Notice that the largest thermopower could be achieved by having a weak coupling to the electrodes (large tunnel barriers) and a molecular resonance close to the Fermi level (large derivative). Since there is no upper bound for $ZT$, very large values can be formally attained within these model calculations. 
	}
	\label{fig:toy}
	\end{center}
\end{figure}

\paragraph{Effective model Hamiltonian$-$} We will first illustrate our approach within a minimal model Hamiltonian in order to highlight different factors influencing  the thermopower of the junction. We consider two electronic levels, which mimic the frontier orbitals of a molecule and which are coupled to semiconducting electrodes as shown in Fig.~\ref{fig:toy}(a). The thermopower $S$  and the figure of merit $ZT$~\cite{comm1} are calculated as a function of the relative position of the HOMO to the Fermi level $E_F$, $\Delta=E_{\textrm{HOMO}}-E_{\textrm{F}}$, and for different coupling strengths  to the electrodes. In Fig.~\ref{fig:toy}(b) and (c), room temperature values for $S$ and $ZT$  are shown. We see that it is possible to obtain $S$ values as large as $\sim$2 mV/K when the molecule-electrode coupling is $\sim$10 meV, and the HOMO level is placed 0.5 eV below $E_F$. Even much higher values of $S$ are possible for weaker coupling strengths.  The optimum value of $\Delta$ is related to the operating temperature and it is smaller for lower temperatures~\cite{citeulike:3500670}. Very high thermopower can be achieved with a delta-function shaped conductance peak near to $E_F$ for  bulk systems~\cite{Mahan1996}; for a model molecular system this behaviour has  been  recently demonstrated~\cite{citeulike:3500670}. Formally, the figure of merit has no upper bound, so that arbitrarily large values can be obtained within a model approach. Though for the realistic molecular junctions  we are going to consider farther below the effective electrode-molecule coupling is on average strong ($\sim$100 meV), the combined use of semiconducting electrodes and chemical tuning still allows for thermopower optimization~\cite{comm0}.

 \begin{figure}
 	\begin{center}
 	\includegraphics[scale=0.85]{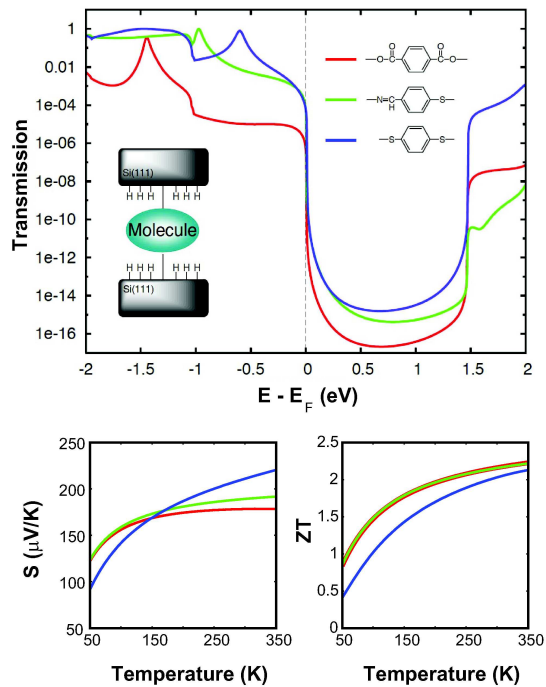}
 	\caption{
 	Top panel: Zero-temperature transmission spectra of a benzene molecule sandwiched between Si electrodes via different linker groups. The  position of the HOMO level is modified due to charge transfer effects. Bottom panel: Temperature dependence of the thermopower $S$ (left) and corresponding figure of merit $ZT$ (right) for the different linker groups.
 	}
 	\label{fig:case1}
 	\end{center}
 \end{figure}

  \begin{figure}
 	\begin{center}
 	\includegraphics[scale=0.85]{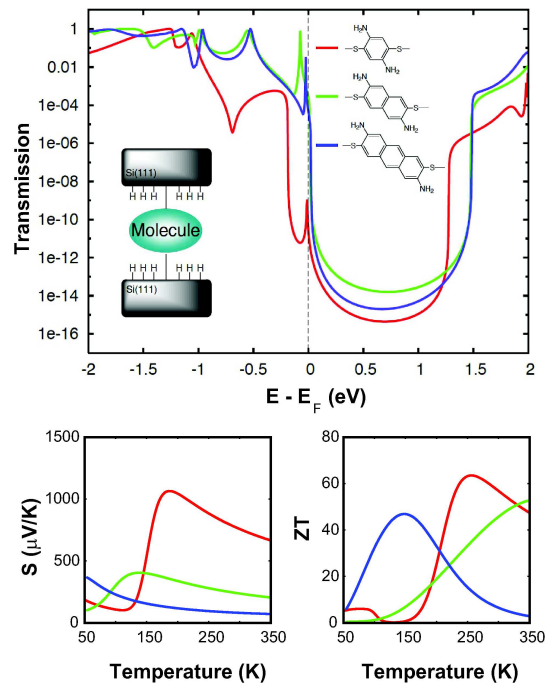}
 	\caption{
 	Top panel: Zero-temperature transmission spectra of three polyacene molecules (benzene, naphtalene, and anthracene) with NH$_2$ side groups and attached to Si electrodes via thiol linkers. For all three molecules the HOMO level lies almost at the corresponding Fermi energy. Notice also that for the benzene molecule the HOMO state moves into the semiconductor band gap. Bottom panel: Temperature dependence of the thermopower $S$ (left) and figure of merit $ZT$ (right).
 	}
 	\label{fig:case3}
 	\end{center}
 \end{figure}

\paragraph{Thermopower of single-molecule junctions$-$} In this section we will discuss  to which extent the analysis presented in the previous paragraphs can be realized in realistic molecular junctions. Methodologically, we will use a density-functional parametrized tight-binding approach (DFTB)~\cite{dftb,gdftb} combined with  Green function techniques to calculate the  transmission function of the junctions and from it the thermoelectric transport properties at zero applied bias. We have addressed two main issues: (i) the role of the linker groups in determining the relative position of the HOMO with respect to the Fermi energy, and (ii) modifications of the thermopower through selective chemical doping of the molecules. 

 Figs. \ref{fig:case1} and \ref{fig:case3} show the transmission spectra of several junctions together with the corresponding thermopower $S$ and figures of merit $ZT$.
From Fig.~\ref{fig:case1}, where the influence of different linker groups on the transmission function around the Fermi level~\cite{comm3} is shown, we  see that thiol linkers turn out to be the most effective linkers. They induce a high-transmission HOMO resonance around 0.5 eV below the Fermi level. Notice also the sharp suppression of the transmission tails above the conduction band edge due to the absence of spectral weight in the semiconducting electrodes: this effect can additionally lead to a large derivative at the Fermi level and to a dramatic increase in $S$ when comparing with metallic electrodes. The other  linkers studied here induce larger energy shifts of the transmission resonances away from the Fermi level and thus seem to be less appropriate for the realization of an efficient thermoelectric molecular junction. Taking this into account we have then investigated small polyacene molecules (benzene, napthtalene, anthracene) with NH$_{2}$ functional groups added at the ortho-positions of the benzene rings, see Fig.~\ref{fig:case3}. As recently shown~\cite{citeulike:3929829}, the controlled torsional motion of side groups can help to tune the thermopower of a molecular junction; this would require however very specific mechanical manipulations at the single molecule level. In the present study we demonstrate a different alternative  route to potentially achieve high thermoelectric efficiency via pure chemical tuning of transmission resonances in combination with non-metallic electrodes. As it comes out from the analysis of our results, the side groups act as a chemical gate which shifts the transmission spectrum towards the Fermi level. As a result, for the three polyacene species studied the HOMO levels are essentially lying at or slightly below  E$_{\textrm{F}}$. Moreover, due to a low weight onto the linker region, the coupling of these levels to the electrodes is rather weak, leading to a strong narrowing. Notice that in contrast to naphtalene and anthracene, the HOMO resonance of the benzene dithiol molecule has a very low transmission; this is related to the fact that the energy shift induced by the side groups moves this level into the semiconducting band gap, where no spectral support exist in the electrodes. Taking into account the narrowness of the HOMO levels  together with their energetic position, we may expect a dramatic increase of the thermopower  when comparing with the junctions studied \eg, in Fig.~\ref{fig:case1}; this is the case as shown in the lower panel of Fig.~\ref{fig:case3}, where maximum $S$-values ranging from 400$-$1000 $\mu V/K$ could be achieved. Also very large figures of merit of up to 60 at about 270 K might be ideally reached. The fact that the maximum values of $S$ and $ZT$ do not necessarily lie at the same temperature demonstrates that the optimization of the thermopower alone is not enough to achieve optimal figures of merit. We stress here that our main focus are  electronic contributions while the vibrational component of the thermal conductance has been neglected at this stage. Experiments~\cite{Reddy2007,wang2007} as well as theoretical studies~\cite{segal2003} suggest that vibrational contributions in organic molecular systems are of the order of  pW/K above room temperature. We expect that the high mismatch between the silicon surface modes and the local vibrational modes of the molecules will lead to similar orders of magnitude in our case. On the other hand, our calculated electronic thermal conductance for \eg, the naphtalene molecule is about 1 pW/K at 1000 K.  Notice that we can write $ZT=ZT_{\textrm{el}}\, \kappa_{\textrm{el}}/(\kappa_{\textrm{el}}+\kappa_{\textrm{ph}})$. Hence, the inclusion of vibrational contributions would lead to a reduction of the second factor by at most  one order of magnitude. Still, the resulting $ZT$ values ($\sim$ 4-6) would be large  when compared to nanowires~\cite{Boukail2008,Hochbaum2008}.

In summary, we have shown in this study  that thermopower engineering in molecular junctions could be attained by an appropriate combination of chemical gating and the use of semiconducting electrodes. The latter provide a way to eliminate  the LUMO channel from the thermopower,  thus lifting the partial cancelation of hole- and electron-like contributions. Though the obtained very large figures of merit should be considered with caution due to the neglect of vibrational contributions, large $ZT$ values might still be achieved by using molecular junctions. We also expect that our results can be scaled up to estimate the thermoelectric efficiency of self-assembled monolayers if the molecule-molecule interactions are not very strong. 

\paragraph{Acknowledgments$-$} The authors aknowledge fruitful discussions with O. Schmidt and Th. Dienel. This work has been supported by the Volkswagen Foundation,
by the European Union under contract IST-021285-2, and by the Deutsche Forschungsgemeinschaft within the priority program ''Nanostructured Thermoelectrics'' (SPP 1386).  We further acknowledge the Center for Information Services and
High Performance Computing (ZIH) at the Dresden University of Technology for computational resources.


\begin{thebibliography}{10}

\bibitem{PhysRevB.47.12727}
L.~D. Hicks and M.~S. Dresselhaus,  Phys. Rev. B {\bf 47},  12727  (1993).

\bibitem{Boukail2008}
A.~I. Boukai, Y. Bunimovich, J. Tahir-Kheli, J.-K. Yu, W.~A.~Goddard III, and J.~R.
  Heath, Nature
  {\bf 451},  168  (2008).

\bibitem{Hochbaum2008}
A.~I. Hochbaum, R. Chen, R.~D. Delgado {\textit{et al.}},
Nature {\bf 451},  163  (2008).

\bibitem{li:3186}
D. Li, Y. Wu, R. Fan, P. Yang, and A. Majumdar,  Appl. Phys. Lett. {\bf 83},  3186  (2003).

\bibitem{nielsch2008}
J. Lee, S. Farhangfar, J. Lee {\textit{et al.}},
Nanotechnology {\bf 19}, 365701 (2008). 

\bibitem{mori2005}
 T. Mori, J. Appl. Phys. {\bf 97}, 093703 (2005).

\bibitem{MingoN._nl034721i}
N. Mingo, L. Yang, D. Li, and A. Majumdar, Nano Lett. {\bf 3},  1713  (2003).

\bibitem{Mingo2008}
I. Savic, D. A. Stewart, and N. Mingo, Phys. Rev. B, {\bf 78}, 235434 (2008).

\bibitem{galli2008}
T.~T.~M. Vo, A.~J. Williamson, V. Lordi, and G. Galli, Nano Lett. {\bf 8}, 1111 (2008).

\bibitem{kim:045901}
W. Kim, J. Zide, A. Gossard {\textit{et al.}},
  Phys. Rev. Lett. {\bf 96},  045901  (2006).

\bibitem{Lyeo2004}
H.-K. Lyeo, A.~A. Khajetoorians, L. Shi {\textit{et al.}},
Science {\bf 303},  816  (2004).

\bibitem{cfr05}
{\em Introducing Molecular Electronics}, Vol.~680 of {\em Lecture Notes in
  Physics}, edited by G. Cuniberti, G. Fagas, and K.~Richter  (Springer,
  Berlin, 2005).

\bibitem{Reddy2007}
P. Reddy, S.-Y. Jang, R.~A. Segalman, and A. Majumdar,  Science {\bf 315},  1568  (2007).

\bibitem{wang2007}
Z. Wang, J. A. Carter, A. Lagutchev {\textit{et al.}},
Science,  {\bf 317}, 787 (2007).

\bibitem{citeulike:3871696}
M. Paulsson and S. Datta,
  Phys. Rev. B {\bf 67},  241403(R)  (2003).


\bibitem{citeulike:3500670}
P. Murphy, S. Mukerjee, and J. Moore, Phys. Rev. B  {\bf 78}, 161406(R)   (2008).

\bibitem{citeulike:4270003}
F. Pauly, J.~K. Viljas, and J.~C. Cuevas,  Phys. Rev. B {\bf 78},  035315  (2008).

\bibitem{citeulike:3806071}
D. Segal,  Phys. Rev. B {\bf 72}, 165426   (2005).

\bibitem{citeulike:4785902}
G.~U. Sumanasekera, B.~K. Pradhan, H.~E. Romero, K.~W. Adu, and P.~C. Eklund,
   Phys. Rev. Lett. {\bf 89},  166801 (2002).

\bibitem{citeulike:3929829}
C.~M. Finch, V.~M. Garc\'{i}a-Su\'{a}rez, and C.~J. Lambert,  Phys. Rev. B {\bf 79}, 033405   (2009).

\bibitem{baranger:nanolett2009}
S.-H. Ke, W. Yang, S. Curtarolo, and H.~U. Baranger,  Nano Lett. {\bf 9}, 1011  (2009).

\bibitem{PhysRevB.70.195107}
J. Koch, F. von Oppen, Y. Oreg, and E. Sela,  Phys. Rev. B {\bf 70},  195107  (2004).

\bibitem{esfarjani:prb2006}
K. Esfarjani, M. Zebarjadi, and Y. Kawazoe,  Phys. Rev. B {\bf 73},  085406 (2006).

\bibitem{muller:jchemphys2008}
K.~H. M{\"u}ller, J. Chem. Phys. {\bf 129},  044708  (2008).

\bibitem{Mahan1996}
G.~D. Mahan and J.~O. Sofo,  Proc.\ Natl.\ Acad.\
  Sci.\ USA {\bf 93},  7436  (1996).

\bibitem{comm2}
The molecules are covalently attached to hydrogen-passivated Si(111) surfaces and relaxed with periodic boundary conditions parallell to the surface~\cite{nozaki2009}. In each unit cell, the top/bottom silicon surfaces  comprise 54 Si atoms forming three Si layers and 8 hydrogen atoms covering the Si(111) surface.  In order to allow for structural relaxation effects of the  surface, the two outermost layers  are included in the relaxation process.

\bibitem{comm1}
Using non-equilibrium thermodynamics, expressions for the Onsager coefficients  in terms of the  transmission function ${\cal{T}}(E)$ can be written~\cite{citeulike:3871696,citeulike:3806071}:
$L_n(T)=\int\limits dE\, (E-E_F)^n (-\partial f(E,T)/\partial E){\cal T}(E)$,
where $f(E,T)$ is the Fermi function. The   Seebeck coefficient $S$ and the electronic part of the  thermal conductance $\kappa_{el}$  can then be obtained as: $S(T)=(-1/eT)(L_1/L_0)$ and $\kappa_{el}(T)=(2/hT)\left(L_2-(L_1^2/L_0)\right)$.
Neglecting  phonon contributions, the figure of merit $ZT$ is $ZT^{-1}=(L_0L_2/L_1^2)-1$. 
The transmission function  ${\cal{T}}(E)={\textrm{Tr}}[G^{r}\Gamma^{\textrm{L}}G^{a}\Gamma^{\textrm{R}}]$ can be computed as a function of the charge injection energy $E$. The broadening functions $\Gamma^{\alpha}$ ($\alpha$=L,R) of the left (L) and righ (R) electrodes are calculated via the self-energies $\Sigma^{r,\alpha}$ as $\Gamma^\alpha=\ii\left(\Sigma^{r,\alpha}-\Sigma^{a,\alpha} \right)$. The retarded molecular Green function is given by
$G^{r}(E)^{-1}=\left[(E+\textrm{i}0^+)I-H_{\textrm{m}}-\Sigma^{r,\textrm{L}}-\Sigma^{r,\textrm{R}}\right]$, $I$ being the identity matrix, and $H_{\textrm{m}}$ the Hamiltonian for the free molecule.


\bibitem{comm0}
For the sake of completeness we have also studied the thermopower in the  specific case of weak coupling (Coulomb blockade) to the electrodes by using the Anderson Hamiltonian to include Coulomb interactions on the two-sites molecule.  We have used nonequilibrium Green functions to calculate the transport properties in this regime~\cite{MCGpaper_104}.
For $U\neq 0$, additional states at energies $\sim\epsilon_{\textrm{HOMO,LUMO}}+U$ emerge; such states could be additionally tuned, thus eventually leading to a further increase of the thermopower. This however requires a separate study, see \eg, Ref.~\cite{citeulike:3500670}.

\bibitem{MCGpaper_104}
B. Song, D.~A. Ryndyk, and G. Cuniberti, Phys. Rev. B {\bf 76},  045408  (2007).

\bibitem{dftb}
T. Frauenheim, G. Seifert, M. Elstner, Z. Hajnal,  {\textit{et~al.}}, phys. stat. sol.(b) {\bf 217},  1  (2000).

\bibitem{gdftb}
A.~Pecchia and A.~Di Carlo, Rep. Progs. Phys. {\textbf{67}}, 1497 (2004).



\bibitem{nozaki2009}
D. Nozaki and G. Cuniberti, Nano Res., to appear (2009).

\bibitem{comm3}
The position of the Fermi energy in our calculations is provided by a self-consistent calculation under the constraint of global charge neutrality in the electrode-molecule system. A pinning of E$_{\textrm{F}}$ at the valence band edge could be achieved {\textit{e.g.}}, by $p$-doping.

\bibitem{segal2003}
D. Segal, A. Nitzan, and P. H{\"a}nggi, J. Chem. Phys. {\bf 119}, 6840 (2003).

\end{thebibliography}
\end{document}